\documentclass{WileyMSP-template}

\usepackage{graphicx}
\usepackage{gensymb}
\usepackage{amsmath}
\usepackage{xcolor}

\definecolor{new}{RGB}{50, 50, 250}

\begin{document}

\pagestyle{fancy}
\rhead{\includegraphics[width=2.5cm]{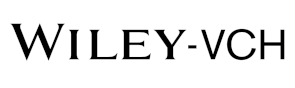}}

\title{Propagation of laser-generated GHz surface acoustic wavepackets in FeRh/MgO(001) below and above the antiferromagnetic– ferromagnetic phase transition}

\maketitle


\author{Iaroslav A. Mogunov*}
\author{Andrey Yu. Klokov}
\author{Nikolay Yu. Frolov}
\author{Andrey I. Sharkov}
\author{Andrey V. Protasov}
\author{Gleb E. Zhezlyaev}
\author{Denis I. Devyaterikov}
\author{Vladimir I. Zverev}
\author{Alexandra M. Kalashnikova}




\begin{affiliations}
Dr. Ia.~A.~Mogunov, Dr. A.~M.~Kalashnikova\\
Ioffe Institute, 194021 St. Petersburg, Russia\\
mogunov@mail.ioffe.ru

Dr. A.~Yu.~Klokov, N.~Yu.~Frolov, Dr. A.~I.~Sharkov\\
P.N. Lebedev Physical Institute of the RAS, 119991 Moscow, Russia

Dr. A.~V.~Protasov, G.~E.~Zhezlyaev, Dr. D.~I.~Devyaterikov\\
Institute of Metal Physics of the Ural Branch of the RAS, 620108 Ekaterinburg, Russia

Dr. V.~I.~Zverev\\
Lomonosov Moscow State University, 119991 Moscow, Russia\\
\textcolor{new}{Sarov branch, Lomonosov Moscow State University, 607189 Sarov, Russia}

\end{affiliations}


\keywords{Surface acoustic wave, photoinduced phase transition, AFM-FM phase transition, FeRh/MgO, interferometry, acoustic dispersion, phase and group velocity}

\begin{abstract}

Magnetoacoustic devices that harness the strong coupling between acoustic waves and magnons have emerged as a promising platform for energy-efficient spintronics.
Laser-generated pulsed surface acoustic waves (SAWs) are particularly attractive for such applications, offering broadband frequency content up to the gigahertz (GHz) range, remote excitation without lithographic patterning, and surface localization for efficient on-chip integration.
This work presents a comprehensive experimental study of laser-generated SAW pulses in the Fe$_{49}$Rh$_{51}$/MgO(001) system.
A thin film of the near-equiatomic FeRh alloy serves both as an opto-acoustic transducer and as a mechanical load that modulates SAW propagation.
The antiferromagnetic to ferromagnetic phase transition in FeRh, occurring slightly above room temperature, is accompanied by abrupt changes in its elastic properties, enabling controlled modification of the SAW excitation efficiency and dispersion characteristics by tuning the sample temperature and laser fluence.
Using 160~fs laser pulses for excitation and time-resolved Sagnac interferometry for detection, the key SAW parameters are evaluated, including amplitude, spectral content, phase and group velocities, and their in-plane anisotropy.
Particular emphasis is placed on the dispersion relation and its anisotropy, which govern the coherent interaction between phonons and magnons and are determined primarily by the FeRh film.

\end{abstract}


\section{Introduction}

The near-equiatomic FeRh alloy has attracted notable interest in spintronics due to its unique first-order phase transition from an antiferromagnetic (AFM) to a ferromagnetic (FM) state slightly above room temperature \cite{FeRh_content_diagram}.
This transition, occurring at 360-390~K in high-quality Fe$_{49}$Rh$_{51}$ thin films \cite{Maat_2005,KOMLEV_2021}, is accompanied by the occurrence of a Rh magnetic moment, 1~\% isostructural expansion of its cubic lattice, and electrical resistivity change \cite{SAW_FeRh_transition,Mag_atoms_1963,ZVEREV_review}, making FeRh a compelling platform for applications in data storage \cite{Marti_memory_2014,Moriyama_memory_2015,Wu_memory_2022}, antiferromagnetic spintronics \cite{Feng_FeRh_review,Fina_FeRh_review}, thermally assisted magnetic recording \cite{Thiele_HAMR_2003}, and neuromorphic computing \cite{FeRh_SAW_neuro_2025}.
FeRh films were epitaxially grown on a variety of substrates such as MgO, Al$_2$O$_3$, SrTiO$_3$, LaAlO$_3$, KTaO$_3$, among which MgO(001) results in FeRh films with exceptional crystalline quality \cite{Xie_substr,Arregi_growth_substrates,Ceballos_substr} making the FeRh/MgO(001) system a basic choice for both fundamental and device applications.
The ability to manipulate the phase transition by temperature, magnetic field \cite{Maat_2005}, or ultrashort laser pulses \cite{Ju_1st_las_ind_PT_2004,Thiele_dyn_2004} further enhances the versatility of this material system.

Surface acoustic waves (SAWs) have emerged as a powerful tool for controlling magnetization in nanostructures: SAWs easily propagate over hundreds of microns \cite{Sugawara_ripples_SAW_2002,Frolov_dispersion}, offer exceptional efficiency in generating and manipulating spin waves via magnetoelastic coupling \cite{Bukharaev_straintronics_2018,Delsing_SAW_roadmap_2019,Vlasov_review_2022}, and enable non-local magnetic control without the need for complex on-chip wiring \cite{Yaremkevich_guided_2021}.
The coherent interaction between phonons and magnons enables a range of functionalities, including nonreciprocal acoustic transport \cite{nonrec_magnetoac_2026,nonrec_SAW_due_to_spin}, spin-wave excitation \cite{Scherbakov_1st_2010}, and even deterministic magnetization switching \cite{Vlasov_SAW_switch}.
Importantly, short pulses of GHz SAWs can be generated optically by focusing femtosecond laser pulses onto the sample surface \cite{Sugawara_ripples_SAW_2002}.
Laser-generated SAWs and spin waves \cite{Kampen_optical_spin_waves_2002} are particularly attractive for all-optical device concepts, as they provide broadband frequency content, complex wavefronts, and the possibility of remote excitation without lithographic patterning \cite{Yaremkevich_guided_2021}.
In order to efficiently design a SAW-based magnetoacoustic device, the parameters of the acoustic waves have to be taken into consideration.

In FeRh/MgO system, the abrupt change in elastic and thermal properties accompanying the phase transition offers a unique opportunity to actively modulate properties of the laser-generated SAWs by temperature or laser fluence.
Recently, the change in the SAW velocities and attenuation across the phase transition was examined for electrically generated continuous-wave SAWs \cite{FeRh_SAW_2026} and thermal SAWs \cite{Ourdani_2024}.
Conversely, the SAWs can induce local phase switching \cite{SAW_FeRh_transition}, and modify magnetic properties \cite{SAW_in_FeRh_FMR}.
In our previous work we demonstrated the amplitude tunability of laser-generated GHz SAWs in FeRh/MgO across the metamagnetic transition \cite{Previous}, however, information on how the phase transition affects the main SAW properties -- dispersion, velocities, and spectrum -- remains incomplete.

SAWs with larger amplitudes can affect \cite{Mogunov_affect} and even induce phase transitions \cite{SAW_FeRh_transition}.
The spectral content of the SAW pulse defines the length or duration of the pulse which is important for clear separation of distinct pulses, as well as information encoding capabilities \cite{pulses_info_spectral}.
Phase velocity is important for coherent acoustic control \cite{ac_diamond_spin_coher,Hashimoto_coher_ac_mag_2018} and group velocity is important for acoustical delays \cite{SAW_group_delay_device}.
SAW velocities also define the type of magnetoalatic interaction \cite{gerevenkov2025}.
The dispersion relation is of utmost importance for utilizing the spin-phonon coupling \cite{Bukharaev_straintronics_2018,Vlasov_review_2022,Li_SAW_mag_review_2021,Yang_SAW_mag_review_2021}.
In the recent works by Ourdani et al. \cite{Ourdani_2024,ourdani_phon_mag_res_BLS} the SAW dispersion was measured in FeRh/MgO by Brillouin light scattering \textcolor{new}{(BLS)}, which is sensitive to relatively large wave vectors $k>$=5~rad/$\mu$m.
Such wavevectors are hardly achievable with laser generation demanding the use of plasmonic resonance \cite{Temnov_acmagplas}, superlattices \cite{Maznev_THz_phon_MQW} or ultraviolet light \cite{extreme_uv_grating}.

In this study, we investigate short pulses of laser-generated SAWs optically generated in epitaxial FeRh/MgO(001) thin films using time-resolved interferometry.
By systematically varying the initial sample temperature and pump laser fluence, we probe the SAW-related response both below and above the AFM–FM transition temperature.
We evaluate amplitudes, spectral content, anisotropies of phase and group velocities, and the dispersion relations of the generated SAW pulses, and compare them with numerical calculations for the two magnetic phases of FeRh.
Our results provide a detailed characterization of SAW propagation in this prospective magnetostructural system and establish a foundation for designing acoustically controlled spintronic devices based on FeRh.

\section{Methods}

\subsection{Sample}

\begin{figure}
    \centering
        \includegraphics[width=0.5\textwidth]{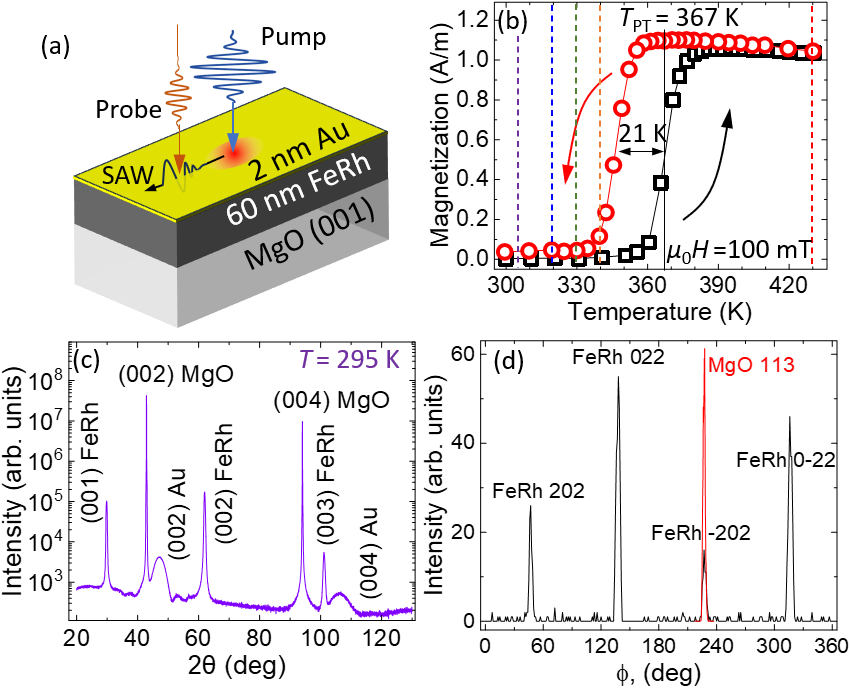}
    \caption{Characterization of the epitaxial Fe$_{49}$Rh$_{51}$ thin film.
    (a)~Schematic illustration of the sample geometry, depicting the pump and probe pulses and the generated SAW pulse;
    (b)~Thermal magnetization hysteresis loop under an in-plane magnetic field $\mu_0H$=100~mT.
    Vertical dashed lines mark the used temperatures;
    (c)~X-ray diffraction $\theta-2\theta$ and (d)~$\phi$ scans acquired at room temperature.
    }
    \label{fig_char}
\end{figure}

We used an epitaxial Fe$_{49}$Rh$_{51}$ thin film with a thickness of 60~nm, capped by a 2~nm Au layer, as illustrated in \textbf{Figure~\ref{fig_char}}(a).
The film was grown on a monocrystalline MgO(001) substrate by molecular beam epitaxy; detailed growth conditions are described in Ref.~\cite{KOMLEV_2021}.
X-ray reflectometry measurements revealed root-mean-square roughness values of $<$0.3~nm for the MgO/FeRh and FeRh/Au interfaces, and 0.7~nm for the Au/air interface.

The FeRh layer adopts a cubic CsCl-type crystal structure, with an epitaxial relationship of FeRh[001]$\|$MgO[001] and FeRh[100]$\|$MgO[110], as confirmed by X-ray diffraction (see Figure~\ref{fig_char}(c)).
At room temperature, the lattice parameters were determined to be $a$=2.980~\AA (in-plane) and $c$=2.992~\AA (out-of-plane), indicating a slight tetragonal distortion arising from epitaxial strain.
With reference to the bulk Fe$_{49}$Rh$_{51}$ lattice constant at room temperature, $a_{\text{bulk}}$=2.991~\AA \cite{Fe49Rh51_cooling_2020}, the resulting epitaxial strains are -0.37~\% for the in-plane parameter and +0.03~\% for the out-of-plane parameter.
The FeRh film grew in a monocrystalline form, as confirmed by in-plane X-ray diffraction $\phi$-scan measurements, shown in Figure~\ref{fig_char}(d).

The AFM-to-FM phase transition was characterized by vibrating sample magnetometry.
Temperature-dependent magnetization curves measured under an in-plane magnetic field $\mu_0$H=100~mT applied along the FeRh[110]$\|$MgO[100] direction are presented in Figure~\ref{fig_char}(b).
This field magnitude was selected to exceed the saturation field of 25~mT in the FM phase \cite{Previous}.
The AFM-FM transition temperature was determined to be $T_{PT}$=367~K, with a thermal hysteresis width of 21~K.

\subsection{Experiment}

\begin{figure}[ht]
    \centering
        \includegraphics[width=0.5\textwidth]{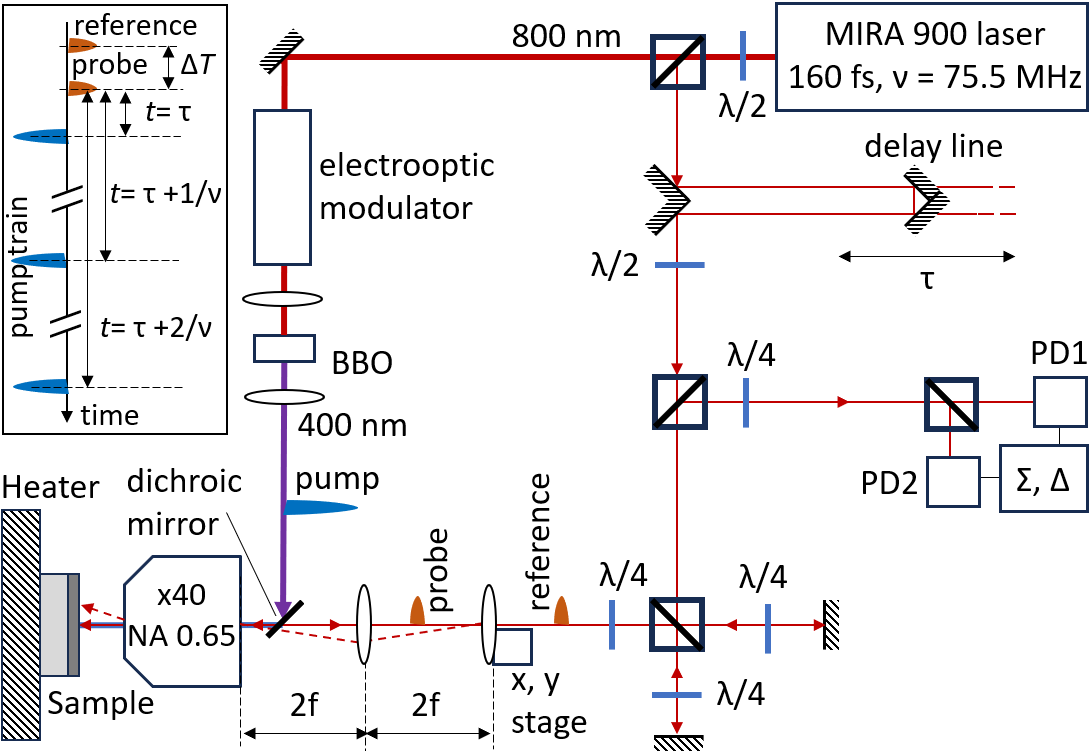}
    \caption{Optical schematic of the Sagnac-type interferometric setup.
    Key components include half-wave ($\lambda/2$) and quarter-wave ($\lambda/4$) plates; BBO, a beta-barium borate crystal for frequency doubling; and PD1 and PD2, the inputs to two photodetectors.
    A 4$f$-scheme is employed for spatial scanning of the probe and reference beams relative to the pump.
    The inset shows time delays involved in experiments.
    }
    \label{fig_scheme}
\end{figure}

The primary experiments were conducted using a laser interferometry technique, a well-established method for the optical detection of laser-generated surface acoustic waves \cite{Sugawara_ripples_SAW_2002,Wright_SAW_vis_2005,Klokov_diamond_2021}.
The experimental configuration, illustrated in \textbf{Figure~\ref{fig_scheme}}, is based on a Sagnac interferometer \cite{Tachizaki_Sagnac_2006}.
This measurement technique exploits a pump-probe method with the probe additionally split into two beams, hereafter referred to as the 'probe' and 'reference' beams, that subsequently interfere at the detector.
One of the interferometer mirrors is formed by the sample surface, enabling the detection of out-of-plane surface displacement $u$ along the surface normal via measurement of the phase shift $\varphi$ between the probe and reference pulses.
This phase shift is detected as $\varphi=\text{Im}[(\Delta R(t)-\Delta R(t-\Delta T))/R]$, where $\Delta T$=500~ps denotes the time delay between the probe and the reference beams.
The scheme also allows the detection of the reflectivity change $\Delta R/R=\text{Re}[(\Delta R(t)+\Delta R(t-\Delta T))/R]$.
The phase shift and the reflectivity change are related to the imaginary and real parts of the complex amplitude reflectivity coefficient $r$ as \cite{MATSUDA_review_2015}:

\begin{equation}\label{eq_refl}
    \varphi=\text{Im}(\frac{\Delta r}{r})=\text{Im}(\frac{\Delta r_{PE}}{r})+2k_{probe}u(z=0); \qquad \frac{\Delta R}{R}=2\text{Re}(\frac{\Delta r}{r})=\text{Re}(\frac{\Delta r_{PE}}{r})
\end{equation}

\noindent where $k_{probe}$ is the wavevector of the probe beam and $\Delta r_{PE}$ is the change in reflectivity due to the photoelastic effect.
The photoelastic contribution is expected to be negligible compared to $u(z=0)$, as FeRh exhibits no spectral features at the probe wavelength of 800~nm \cite{Saidl_opt_params_2016,Rhee_opt_1995}, suggesting minimal photoelastic response at this wavelength.
This technique allows detection of quasi-Rayleigh waves but is not sensitive to Love waves, as the latter have no displacement along the surface normal \cite{FARNELL197235}.

A schematic of the optical setup is shown in Figure~\ref{fig_scheme}.
The light source is a Ti:Sa Coherent MIRA~900 laser, producing 160~fs pulses at a central wavelength of 800~nm and a repetition rate of $\nu$=75.5~MHz, corresponding to a temporal spacing of $1/\nu$=13~ns between consecutive pulses.
The laser beam is split into two paths.
The first serves as the pump beam, which is modulated at 548~kHz using an electro-optic modulator, frequency-doubled in a beta-barium borate (BBO) crystal, and directed via a dichroic mirror into a x40 microobjective with numerical aperture (NA) of 0.65.
The pump beam is linearly polarized and focused onto the sample surface to a circular spot with a full width at half-maximum (FWHM) of approximately 1.3~$\mu$m.
The second beam path is directed through a mechanical delay line producing the time delay $\tau$ with resolution of $\sim$0.2~ps.
This beam is subsequently split into the probe and reference beams within the interferometer.
Both of them travel identical optical paths, but arrive on the sample surface with a time delay $\Delta T$=500~ps, determined by the length of the interferometer.
The probe and reference beams are circularly polarized and pass through a 4$f$-scheme with a movable lens.
The beams are then focused by a microobjective into circular spots with FWHM of approximately 2~$\mu$m.

By translating the lens using a motorized stage, the probe and reference beams can be scanned across the sample surface relative to the pump beam with a spatial resolution better than 1~$\mu$m.
The pump, probe and reference beams arrive at the sample surface with distinct time delays, shown in the inset of Figure~\ref{fig_scheme}.
For the probe beam chosen within the laser pulse train, we detect responses from several pump beams from the same train when the probe position is varied on the sample surface.
These delays are $t$=$\tau$, defined by the delay line, and $t$=$\tau+n/\nu$ ($n$=1, 2) defined by both the time delay and the laser repetition rate.
For example, for $\tau$=3~ns, the time delay values are $t$=3, 16 and 29~ns.

The probe and reference beams reflected from the sample retrace the interferometer path, where they recombine and interfere temporally before being directed onto two photodetectors.
In experiments, either the difference or the sum of the detector channels is recorded, providing signals proportional to the imaginary part (phase $\varphi$) and the real part (amplitude, $\Delta R/R$) of reflectivity change, respectively.
The sample is mounted on a resistive heater, enabling thermally induced phase transition in FeRh with a temperature stability of 1~K.
All measurements were performed in the absence of an externally applied magnetic field.

\section{Results}

Experiments were carried out at initial sample temperatures $T_0$=305, 320, 330, 340 and 430~K marked in Figure~\ref{fig_char}(b).
The temperature of 430~K exceeds the phase transition temperature $T_{PT}$, while the remaining temperatures lie below $T_{PT}$.
The temperatures lying inside the AFM-FM phase coexistence range in the thermal hysteresis were not accessible, as both excitation and detection utilized laser pulse trains, and the first laser pulse would have permanently altered the FeRh phase within this temperature interval.
The pump laser fluence $W$ was varied up to a maximum of 30~mJ/cm$^2$ limited by the onset of laser-induced damage to the film.
SAWs are generated within the FeRh film following absorption of the pump laser pulse.
The optical penetration depth at the pump wavelength is 19~nm in the antiferromagnetic (AFM) phase \cite{Rhee_opt_1995}, which is less than the film thickness of $h$=60~nm.
The MgO substrate remains transparent at both the pump and probe wavelengths.

\begin{figure}[ht]
    \centering
        \includegraphics[width=0.7\textwidth]{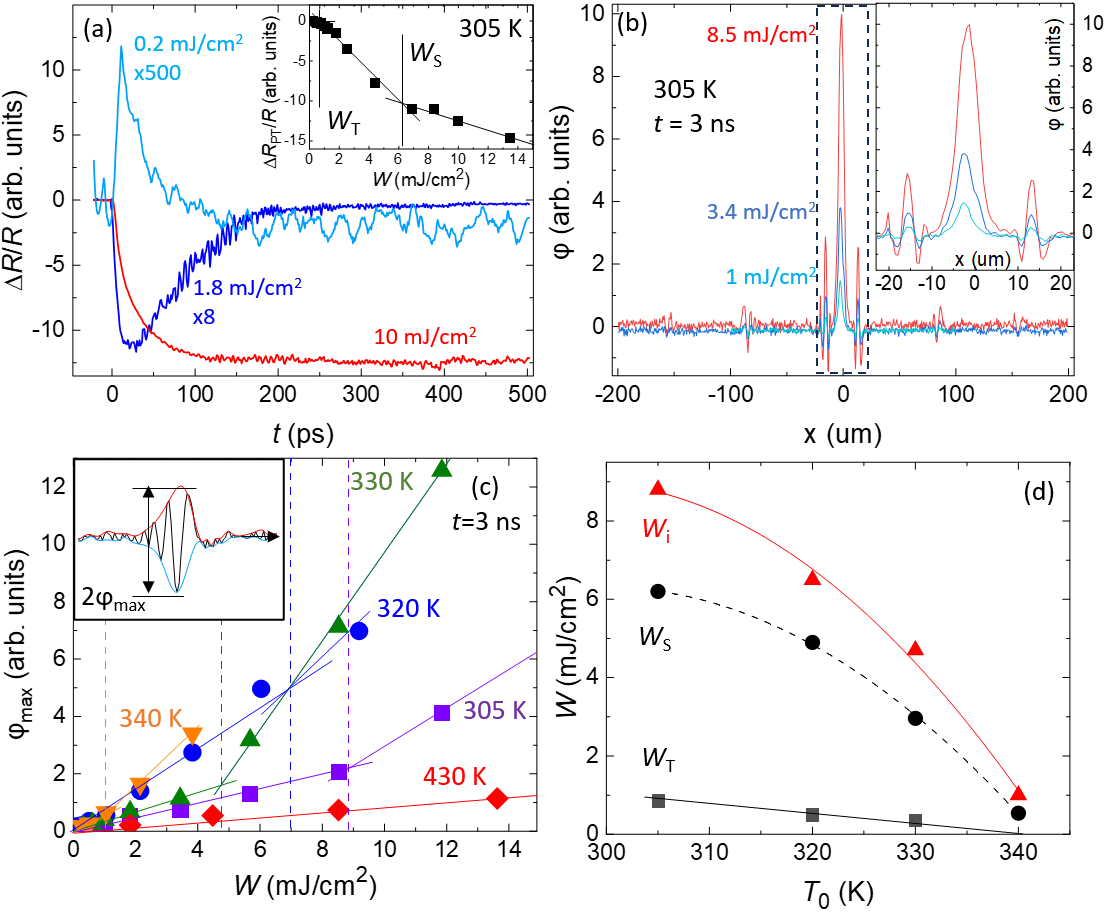}
    \caption{Experimental 1D scans.
    (a)~Time-resolved transient reflectivity change $\Delta R(t)/R$ measured at $T_0$=305~K for various pump fluences $W$ (indicated in the figure) with subtracted thermal background.
    Inset: fluence dependence of the amplitude of the phase-transition-related reflectivity change $\Delta R_{PT}/R$ at the same temperature, with the threshold fluence $W_T$ and saturation fluence $W_S$ marked;
    (b)~Phase shift $\varphi$ as a function of the probe beam position $x$ at $T_0$=305~K, recorded at a fixed time delay $t$=3~ns for several pump fluences.
    The inset shows a magnified view of the region near $x$=0;
    (c)~Pump fluence dependence of the SAW envelope amplitude (see inset for definition) measured at a time delay $t$=3~ns for several initial temperatures $T_0$.
    Vertical dashed lines indicate the inflection points $W_i$ for temperatures below $T_{PT}$;
    (d)~Temperature dependence of the threshold fluence $W_T$, saturation fluence $W_S$ and inflection fluence $W_i$.
    }
    \label{fig_signals}
\end{figure}

For each temperature below $T_{PT}$, we first perform time-resolved measurements of the relative reflectivity change $\Delta R(t)/R$ as a function of time delay $t$ at a fixed spatial position $x=y=0$ for a series of pump fluences, as shown in \textbf{Figure~\ref{fig_signals}}(a).
Here, $x\|[100]_{MgO}$ and $y\|[010]_{MgO}$ directions on the sample surface.
The goal of this procedure is to find the threshold $W_T$ and saturation $W_S$ fluences of the photoinduced phase transition at the temperatures used, which govern the SAW generation mechanisms \cite{Previous}.
To isolate the component associated with the photoinduced phase transition $\Delta R_{PT}/R$, the signal obtained at a subthreshold fluence of 0.2~mJ/cm$^2$($<W_T$) was scaled and subtracted from the data as in Refs.~\cite{Previous,Mattern_8ps,Bergman_2006}.
The thermal background was also subtracted for each curve.
The inset in Figure~\ref{fig_signals}(a) shows a fluence dependence of $\Delta R_{PT}/R$ at $T_0$=305~K.
This dependence exhibits three distinct linear regimes bounded by the threshold fluence $W_T$ and the saturation fluence $W_S$.
For $W<W_T$, no phase transition is induced, and the sole mechanism of SAW generation is thermoelasticity commonly considered for metal-on-substrate systems \cite{Xu_las_SAW_model}.
In the intermediate regime $W_T<W<W_S$, a partial phase transition occurs, resulting in a transient mixed-phase state.
For fluences above the threshold, the SAW generation is governed by both thermoelasticity and phase-transition-related lattice expansion, the latter reaching its maximal potency at $W=W_S$ \cite{Previous}.
For $W>W_S$, the entire probed volume of the FeRh film undergoes transformation into the FM phase.
The extracted values of $W_T$ and $W_S$ are presented as a function of temperature in Figure~\ref{fig_signals}(d).
These fluences correspond well to previous reports for thin FeRh films \cite{Previous,Mattern_8ps,Bergman_2006}.
For $T_0>T_{PT}$, photoinduced phase transition cannot be triggered, and $\Delta R_{PT}$=0.

Following determination of $W_T$ and $W_S$ at each temperature, we measured the phase shift $\varphi$ between the probe and reference beams as a function of the spatial coordinates $x$ and $y$ of the probe pulse on the sample surface, with the time delay fixed at $t$=3~ns.
Representative line scans along the $x$ direction for several fluence values are shown in Figure~\ref{fig_signals}(b).
Asymmetry between the $x>0$ and $x<0$ regions is attributed to a measurement artifact.
The inset in Figure~\ref{fig_signals}(b) confirms that at a time delay of 3~ns, the SAW pulse has already propagated away from the pump-excited area around $x$=0 where $\varphi$ is nonzero.
Therefore, the observed SAW pulses propagate in the unperturbed sample at the initial temperature $T_0$.
Additional wave packets observed at $x=\pm80~\mu$m and $\pm160\mu$m originate from preceding pump pulses within the laser pulse train, corresponding to propagation times of 16 and 29~ns, respectively.
The observed wave packets were identified as quasi-Rayleigh SAW pulses in the FeRh/MgO(001) system.
Their propagation velocity $\sim$5480~km/s, which will be discussed in detail below, closely matches the Rayleigh wave velocity in bare MgO(001) along the [100] direction (5500~m/s) \cite{Lee_MgO_SAW_1995}.
We note that at relatively low frequencies the SAW velocity is predominantly governed by the substrate, since the acoustic wavelength is large compared to the 60~nm FeRh film thickness \cite{FARNELL197235}, as can be seen in the inset of Figure~\ref{fig_signals}(b).

\begin{figure}
    \centering
        \includegraphics[width=\textwidth]{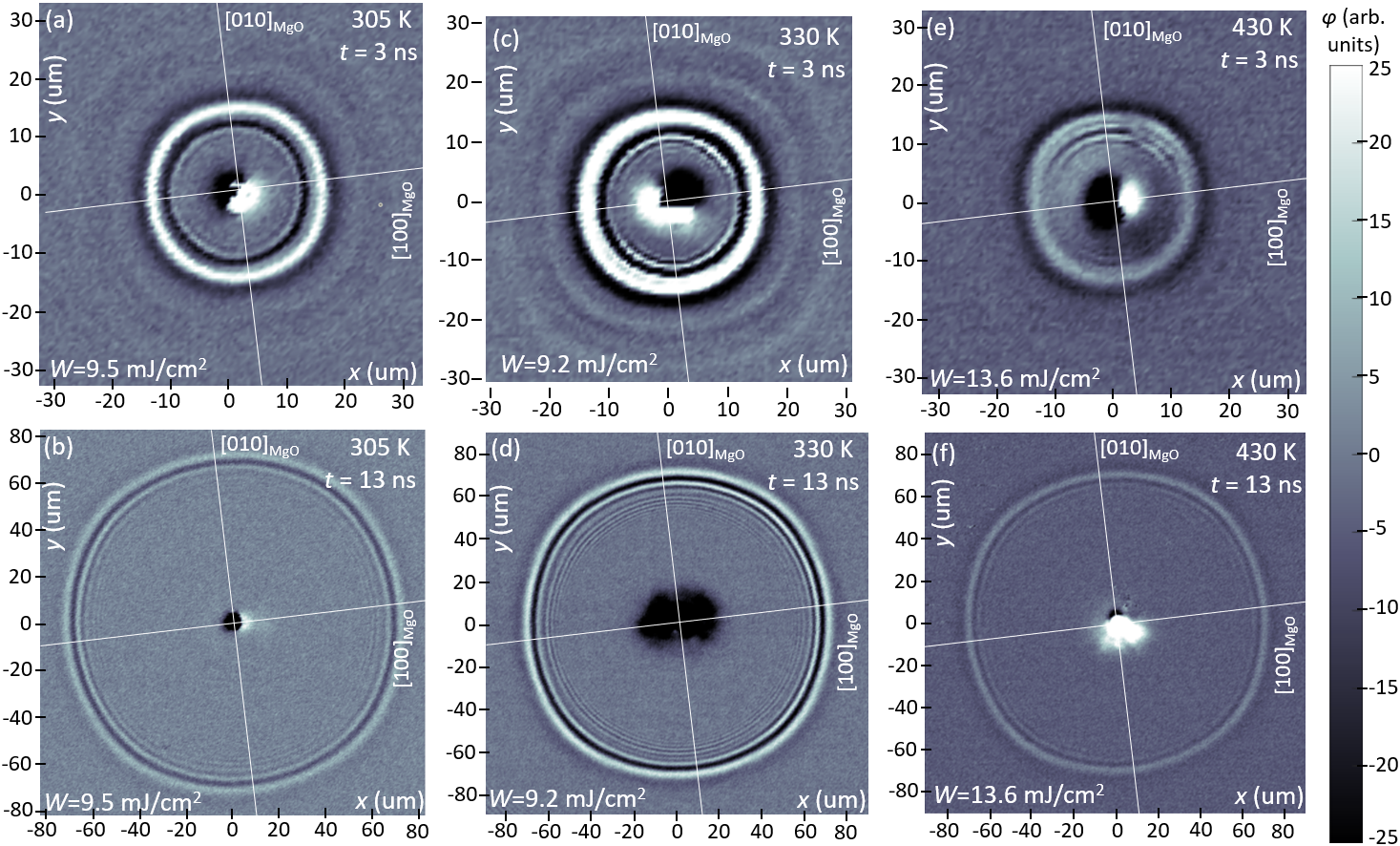}
    \caption{Two-dimensional spatial maps of the measured phase shift $\varphi(x,y)$.
    Data are presented for initial temperatures $T_0$=305~K (a, b), $T_0$=330~K (c, d), and $T_0$=430~K (e, f),
    recorded at time delays $t$=3~ns (a, c, e) and $t$=13~ns (b, d, f).
    Note the different spatial scale for $t$=3 and 13~ns.
    A common color scale applies to all panels.
    }
    \label{fig_2D}
\end{figure}

Two-dimensional (2D) spatial maps of the phase shift $\varphi(x,y)$ were acquired at an above-saturation pump fluence for two temperatures corresponding to the AFM phase (305~K and 330~K) and one corresponding to the FM phase (430~K), as presented in \textbf{Figure~\ref{fig_2D}}.
Maps are shown for two time delays: $t$=13~ns (upper panels) and $t$=3~ns (lower panels).
Note that the spatial scale differs between the image pairs for the same $T_0$.
In all two-dimensional maps, the SAW pulse appears as a distorted circular wavefront, with the distortion arising from the fourfold symmetry of the underlying MgO(001) substrate.
A pronounced reduction in signal amplitude is observed at $T_0$=430~K($>T_{PT}$), consistent with previous observations \cite{Previous}.
Additionally, a weaker, faster-propagating wave is discernible in Figure~\ref{fig_2D}(a) and \ref{fig_2D}(c).
This feature is attributed to a surface-skimming longitudinal acoustic wave \cite{Sugawara_ripples_SAW_2002,Saito_method} and will not be considered further, as it does not constitute a true surface wave and exhibits comparatively low amplitude.

In the following section, we analyze the acquired signals to extract the properties of SAWs propagating in the FeRh/MgO(001) system for initial temperatures both below and above $T_{PT}$.
The extracted parameters include amplitude, spectral content, wavevectors, phase and group velocities, and phase velocity dispersion.
The in-plane anisotropy of the velocity and its dispersion is also examined.

\section{Discussion}

\subsection{Amplitudes and wavenumbers}

We first investigated the temperature and fluence dependence of the measured SAW pulse amplitudes, $\varphi_{max}(W,T_0)$.
The analysis focused on $x$-scan data analogous to those presented in Figure~\ref{fig_signals}(b).
The SAW amplitude was determined by averaging the magnitudes of the upper and lower envelopes, as illustrated in the inset of Figure~\ref{fig_signals}(c).
For the three time delays examined ($t$=3, 16 and 29~ns), the fluence and temperature dependencies exhibited consistent behavior, therefore, the following discussion is focused on results obtained at $t$=3~ns.

Figure~\ref{fig_signals}(c) presents the SAW amplitude as a function of pump fluence for several initial temperatures $T_0$.
In agreement with our previous work \cite{Previous}, for $T_0<T_{PT}$, the amplitude exhibits a nonlinear increase with increasing fluence and, for a fixed fluence, rises monotonically with $T_0<T_{PT}$.
When $T_0$ exceeds $T_{PT}$, the fluence dependence becomes linear and the overall amplitude is substantially reduced.
This behavior is explained by activation of the phase-transition mechanism of the SAW generation when $W$ exceeds the threshold fluence, which adds to the thermoelastic mechanism, present at all temperatures \cite{Previous}.
For $T_0<T_{PT}$, vertical lines in Figure~\ref{fig_signals}(c) denote the inflection points $W_i$ in $\varphi_{max}(W,T_0)$.
These inflections are associated with the characteristics of the photoinduced phase transition in FeRh, specifically threshold or saturation fluences.
The temperature dependence of the inflection fluence $W_i$ is shown in red in Figure~\ref{fig_signals}(d) and exhibits a distinctly nonlinear behavior.
The temperature dependencies of the threshold fluence $W_T$ and saturation fluence $W_S$ are shown as solid and dashed black curves, respectively.
The linear decrease in $W_T$ with increasing temperature is explained by the reduction of the laser energy required to induce the phase transition as $T_0$ gets closer to $T_{PT}$.
The nonlinear temperature dependence of $W_S$ is attributable to cumulative heating effects arising from the high repetition rate of the laser system, combined with the fact that the probe and reference beams together contribute an additional fluence of approximately $\sim$2~mJ/cm$^2$.
Both $W_T(T_0)$ and $W_S(T_0)$ extrapolate to zero at $T_0$=342~K, a temperature notably lower than $T_{PT}$=367~K, further indicating the presence of steady-state heating during the experiment.
From the comparison in Figure~\ref{fig_signals}(d), we conclude that the observed inflection points correspond to saturation fluences, albeit derived from SAW amplitude rather than from the pump-induced reflectivity change.
Such SAW-determined saturation fluence exceeds the optically obtained $W_S$ as it samples the entire FeRh film thickness.
This is because the laser-generated SAW is larger than the 60~nm FeRh film, and its slow generation allows the heat diffusion to redistribute laser energy across the full film thickness \cite{Previous,Mattern_8_and_50ps}.

\begin{figure}
    \centering
        \includegraphics[width=0.7\textwidth]{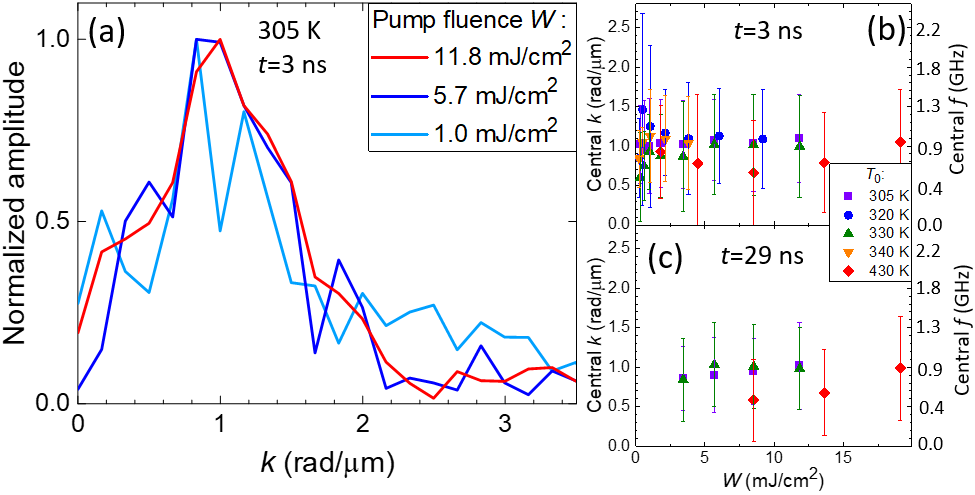}
    \caption{Spectral analysis of the detected SAW pulses.
    (a)~Normalized FFT spectra of the spatial scans $\varphi(x)$ acquired at $T_0$=305~K and $t$=3~ns for a set of pumnp fluences $W$;
    (b, c)~Pump fluence dependence of the central wavenumber $k_x$ and the corresponding phonon frequency for the indicated initial temperatures  $T_0$, shown for propagation times $t$=3~ns (b) and $t$=29~ns (c).
    Error bars denote the FWHM of the spectral distributions.
    }
    \label{fig_fft}
\end{figure}

SAW wavevectors were extracted via fast Fourier transform (FFT) of the spatial scans shown in Figure~\ref{fig_signals}(b).
The FFT was performed along the $x$ direction, yielding the in-plane wavevector component $k_x$.
Normalized FFT spectra of the $\varphi(t)$ signals acquired at $T_0$=305~K and $t$=3~ns are presented in \textbf{Figure~\ref{fig_fft}}(a) for a set of pump fluences.
The dependencies of the central wavenumber on pump fluence and temperature are shown in Figure~\ref{fig_fft}(b) and \ref{fig_fft}(c) for time delays of $t$=3 and $t$=29~ns, respectively.
Data for $t$=29~ns have a significantly lower signal-to-noise ratio due to the attenuation of the SAW upon propagation and are not shown for small fluences.
The central wavenumber of the generated SAWs in these experiments is $k_x$=1~rad/$\mu$m.

\textcolor{new}{Two factors affect the frequency composition of the generated SAWs.
The pump laser spot size limits the in-plane $k$-vectors of the generated phonons, and the generation mechanism limits their frequencies.
In the case of FeRh, we distinguish thermoelastic and phase transition mechanisms \cite{Previous}.
The former is thermal in nature, and the resulting phonon wavenumbers are limited only by the rate of laser-induced temperature rise, and even for 305~K those are an order of magnitude higher than the ones observed in Figure~\ref{fig_fft}, as is evident from the BLS studies \cite{Ourdani_2024}.
The phase transition mechanism has a timescale of the lattice transformation around 100~ps \cite{Quirin_xray_2012}, and the corresponding SAW wavenumber of 11.5~rad/$\mu$m also lays far beyond the observed ones.
Therefore, the laser generated SAW wavenumbers are defined by the pump laser spot size.
The smaller pump size has led to generation of SAWs with larger $k$, as reported in Ref.~\cite{Previous}.}

Analysis of Figure~\ref{fig_fft}(b) and \ref{fig_fft}(c) yields two key observations.
First, the wavevectors exhibit no dependence on pump fluence, indicating that the SAW spectrum is governed by the laser spot size and is independent on the activation of the phase-transition mechanism of SAW generation.
Second, a slight decrease in frequency is observed upon SAW propagation, consistent with frequency-dependent attenuation -- a phenomenon commonly reported in picosecond acoustics \cite{VANCAPEL_review_2015}.
The FWHMs of the spectral distributions shown in Figure~\ref{fig_fft}(a) were found to be independent of both temperature and fluence, with a value of approximately 1.2~rad/$\mu$m.
The values of FWHM serve as error bars in Figure~\ref{fig_fft}(b) and \ref{fig_fft}(c).

\subsection{Phase and group velocities and their anisotropy}

\begin{figure}
    \centering
        \includegraphics[width=\textwidth]{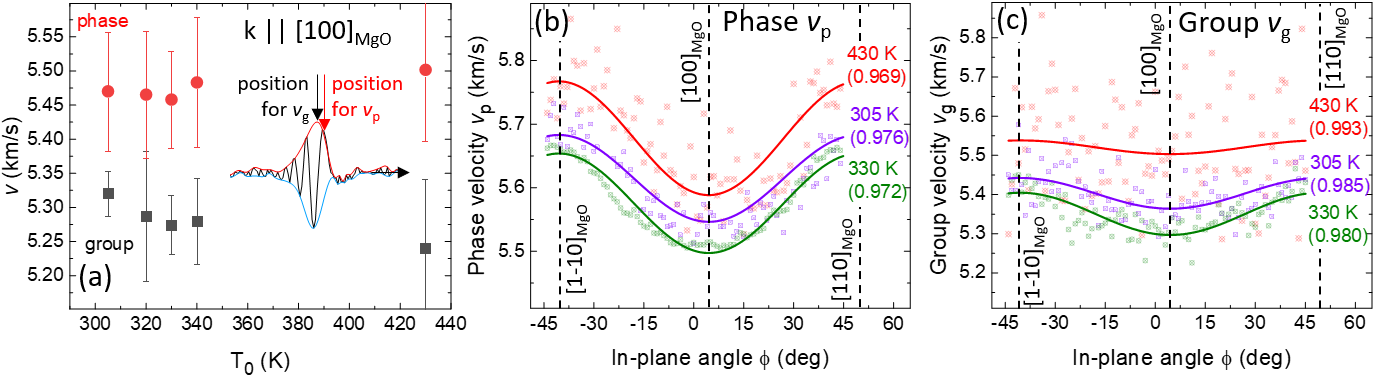}
    \caption{Group and phase velocities of SAWs.
    (a)~Temperature dependence of the phase velocity $v_p$ and group velocity $v_g$ along the MgO[100] direction.
    The inset illustrates the method used to determine $v_p$ and $v_g$ from the experimental waveforms;
    (b-c)~Azimuthal angle dependence of the phase velocity~(b) and group velocity~(c), extracted from the two-dimensional spatial maps shown in Figure~\ref{fig_2D} for several temperatures.
    The ratios $v_{[100]}/v_{[110]}$ are indicated in parentheses.
    The solid lines are fits by a sine function.
    }
    \label{fig_speed}
\end{figure}

Group velocities were determined by tracking the temporal evolution of the envelope maxima, while phase velocities were obtained by tracking the leading maxima of the oscillatory $\varphi$ signal, as illustrated in the inset of \textbf{Figure~\ref{fig_speed}}(a).
Since SAWs propagate through regions of the sample unaffected by the pump pulse, no dependence of velocity on pump fluence $W$ is found.
SAW velocities along the MgO[100] direction were extracted from one-dimensional spatial scans analogous to those shown in Figure~\ref{fig_signals}(b) by comparing signals at distinct time delays $t$=3, 16, and 29~ns.
The resulting group velocities $v_g\sim$5280~m/s and phase velocities $v_p\sim$5480~m/s are presented in Figure~\ref{fig_speed}(a) as a function of the sample temperature.
The extracted phase velocities allow us to calculate the frequencies of acoustic phonons in the SAW pulse using the relation $f=kv_p/2\pi$, the values of $f\sim$1~GHz are shown in Figure~\ref{fig_fft}(b,c).
For all temperatures, the relation $v_g<v_p$ holds, which is expected for SAWs in a system where a thin film mechanically loads a substrate \cite{FARNELL197235}.
The phase and group velocities show negligible temperature dependence.
No abrupt change in velocity is observed across the phase transition, pointing at the dominant role of the MgO substrate in governing SAW velocities.

\textcolor{new}{The obtained SAW parameters depend strongly on the combination of wavenumber (frequency) and FeRh film thickness, $kh$, defining the degree to which the FeRh film affects the SAW propagation, as well as on the material of the substrate.
In our previous work \cite{Previous} on the same sample, we found slightly different values for phase (5.4~km/s) and especially group (4.3~km/s) velocities of the laser generated SAWs due to the smaller pump laser spot size on the sample resulting in higher $k\sim3.5$~rad/$\mu$m.
The frequencies of SAWs in that work were centered at 3.1~GHz.}

The anisotropy of SAW propagation velocities was investigated using the two-dimensional $\varphi(x,y)$ maps presented in Figure~\ref{fig_2D}.
Analysis was performed for two time delays ($t$=3~ns and $t$=13~ns) and three initial temperatures ($T_0$=305, 330, and 430~K).
A clear fourfold symmetry, evident in the spatial maps in Figure~\ref{fig_2D}, is confirmed by the azimuthal angle $\phi$ dependence of both phase and group velocities, shown in Figure~\ref{fig_speed}(b) and \ref{fig_speed}(c), respectively.
This anisotropy originates primarily from the MgO substrate.
However, the epitaxial monocrystalline FeRh film, whose cubic lattice is rotated by 45~deg in-plane relative to MgO \cite{Arregi_growth_substrates}, may alter the velocity anisotropy relative to the bare substrate.
SAW velocities are lower along the MgO[100] direction and higher along the MgO[110] direction, with a ratio of approximately 0.975 for the phase velocity and 0.985 for the group velocity.
For reference, the corresponding ratio for Rayleigh SAWs propagating on a bare MgO(001) substrate is 0.966 \cite{Lee_MgO_SAW_1995}, indicating that the presence of the 60~nm FeRh film exerts a discernible, but small, influence on the velocity anisotropy at \textcolor{new}{acoustic} frequencies near 1~GHz.

\subsection{Dispersion and its anisotropy}

Rayleigh SAWs propagating on a bare substrate are inherently nondispersive \cite{Viktorov_book_SAW}, consequently, any observed dispersion is attributable to the presence of the FeRh film and is characteristic of the FeRh/MgO(001) heterostructure.
Here we evaluate the SAW dispersion within the accessible wavenumber range of $k$=0.4-2~rad/$\mu$m, determined by the spectral content of our SAW pulses (see Figure~\ref{fig_fft}).
Signatures of dispersion are evident in the spatial scans shown in Figure~\ref{fig_signals}(b,c) and \ref{fig_2D}(b,d,f), particularly when comparing pulses emitted at time delays of $t$=3, 16, and 29~ns.
Acoustic dispersion, i.e. the dependence of phase velocity on wavenumber, manifests as a chirped pulse shape \cite{VANCAPEL_review_2015}, characterized by a spatial or temporal variation of the phonon frequency within the pulse, a phenomenon commonly observed in laser-generated SAWs \cite{Sugawara_ripples_SAW_2002,Frolov_dispersion,Previous,WRIGHT_dispersion_2002}.
Thus, information regarding the SAW dispersion is encoded in the shape and spectral composition of the SAW pulse train.

While the dispersion of laser-generated SAWs is typically extracted using two-dimensional FFT analysis \cite{Sugawara_ripples_SAW_2002,WRIGHT_dispersion_2002}, this direct approach is highly sensitive to noise and alignment inaccuracies.
Instead, we employed a method developed in Ref.~\cite{Frolov_dispersion} to reconstruct the phase velocity dispersion $v_p(k)$ of the laser-generated SAW.
This technique assumes an isotropic medium, a reasonable approximation in the present case given that the velocity anisotropy ratio $v_p^{[100]}/v_p^{[110]}$ is close to unity (see Figure~\ref{fig_speed}(b)).
The procedure is as follows.
First, FFT was performed separately on the signals acquired at time delays of $t$=3~ns and $t$=16~ns, yielding oscillatory Fourier-domain signals with smooth envelopes.
The dispersion relation is encoded in the phase of the oscillatory component, expressed as $kv_p(k)t-\theta(k)$, where $\theta(k)$ is the initial phase.
Subsequently, a Hilbert transform was applied to obtain the analytic signal to extract its argument and modulus.
Finally, the arguments of the analytic signals for $t$=3~ns and $t$=16~ns were subtracted from each other, enabling extraction of the phase velocity dispersion $v_p(k)$.

\begin{figure}
    \centering
        \includegraphics[width=0.7\textwidth]{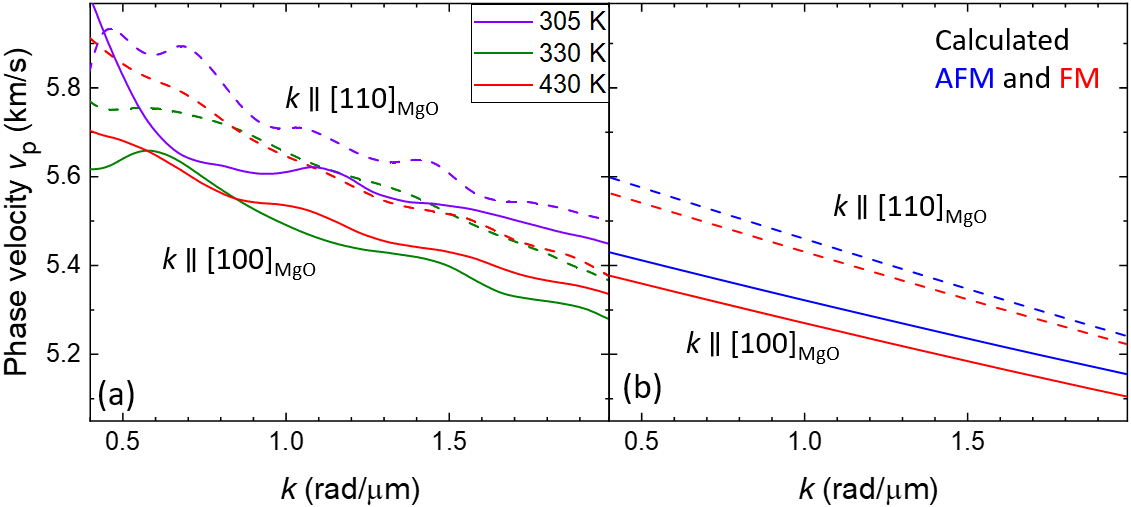}
    \caption{Dispersion of SAWs in the FeRh/MgO(001) system.
    (a)~Phase velocity dispersion $v_p(k)$ extracted from the two-dimensional spatial maps shown in Figure~\ref{fig_2D} along the MgO[100] direction (solid lines) and MgO[110] direction (dashed lines) for several temperatures.
    (b)~Calculated phase velocity dispersions for the AFM phase (blue) and the FM phase (red) along the MgO[100] (solid) and MgO[110] (dashed) directions.
    }
    \label{fig_dispersion}
\end{figure}

This method was applied to the two-dimensional maps shown in Figure~\ref{fig_2D}, considering the two principal crystallographic directions MgO[100] and MgO[110].
The resulting dispersion curves are presented in \textbf{Figure~\ref{fig_dispersion}}(a) for wavevectors along the MgO[100] (solid lines) and MgO[110] (dashed lines) directions for the AFM (violet and green) and FM (red) phases of FeRh.
Additionally, we ensured that the dispersions are independent of $W$.
The nonlinearities of $v_p(k)$ at $k<0.6$~rad/$\mu$m present for some data are an artifact inherent to the reconstruction procedure \cite{Frolov_dispersion}, arising from the segmentation of the original signal into individual pulses and the finite window size employed in this segmentation.

For all signals present in Figure~\ref{fig_dispersion}(a), the phase velocity decreases with increasing wavenumber $k$, accounting for the chirped pulse shapes observed in Figure~\ref{fig_signals}(b,c) and \ref{fig_2D}(b,d,f), where higher-frequency components trail in the pulse tail.
The extracted dispersions do not show a clear dependence on the phase of the FeRh film.
Velocities along the MgO[110] direction are clearly higher than along MgO[100], in agreement with  Figure~\ref{fig_speed}(b) and the known anisotropy of bare MgO(001) \cite{Lee_MgO_SAW_1995}.
The slope $dv_p/dk$ is slightly higher for the wavevectors along MgO[110] (see \textbf{Table~\ref{tab:params}}).

Numerical calculations of SAW dispersion in the FeRh/MgO system were performed by solving the elastic wave propagation equation \cite{FARNELL197235,Viktorov_book_SAW} while accounting for the anisotropy of both the film and the substrate.
In the calculations, we used the known elastic tensor of MgO \cite{Lee_MgO_SAW_1995} and a recently reported elastic tensor for a thin FeRh film \cite{Ourdani_2024}.
Both phases of FeRh were considered, while the temperature dependence of the elastic parameters of MgO was neglected.
The calculated phase velocity dispersion $v_p(k)$ for the lowest quasi-Rayleigh mode is shown in Figure~\ref{fig_dispersion}(b) for wavevectors along the MgO[100] (solid lines) and MgO[110] (dashed lines) directions, for both phases of FeRh.

The calculated phase velocity dispersions agree reasonably well with those extracted from the measured data with differences below 5~\% and successfully capture the dependence of $v_p$ on the crystallographic direction.
At the central wavenumber $k$=1~rad/$\mu$m, the anisotropy ratios obtained from the calculations are $v_p^{[100]}/v_p^{[110]}$=0.975 for the AFM phase and 0.970 for the FM phase, both of which are consistent with the experimentally determined value of approximately 0.975 (see Figure~\ref{fig_speed}(b)).
The slope $dv_p/dk$ is also well reproduced for both crystallographic directions considered.
The calculations predict a higher phase velocity in the AFM phase of FeRh compared to the FM phase, however, this difference is smaller than 0.6~\% making it indescribable for the experimental 75~$\mu$m travel distances of the laser-generated SAW pulses.
The 5~\% discrepancy in values of $v_p$ may originate from the FeRh/MgO interface, which can affect velocity dispersion \cite{Frolov_dispersion,SAW_interface_dispersion}, and is related to epitaxial stresses, which vary upon the phase transition in FeRh \cite{Arregi_growth_substrates,Bordel_mag_aniz_2012}.

\subsection{\textcolor{new}{Summary of SAW parameters}}

A summary of the SAW parameters evaluated in this work is provided in Table~\ref{tab:params}.
\textcolor{new}{To put the obtained parameters of SAWs in FeRh/MgO structure in perspective, we discuss also recent reports on SAWs in FeRh-based structures \cite{SAW_FeRh_transition,FeRh_SAW_neuro_2025,FeRh_SAW_2026,Ourdani_2024,SAW_in_FeRh_FMR,ourdani_phon_mag_res_BLS}.}

\begin{table}[h]
    \caption{Summary of parameters characterizing laser-generated SAW pulses in a 60~nm thick Fe$_{49}$Rh$_{51}$ film grown on MgO(001), as extracted from the experimental data \textcolor{new}{in a wavevector range 0.4-2~rad/$\mu$m.}}
    \centering
        \begin{tabular}[htbp]{l c c c c c}
        \hline
        Property & 305 K & 320 K & 330 K & 340 K & 430 K \\
        \hline
        Phase velocity of the leading oscillation, $v_p$ [km/s] & 5.47$\pm$0.09 & 5.46$\pm$0.10 & 5.46$\pm$0.07 & 5.48$\pm$0.10 & 5.50$\pm$0.11 \\
        Group velocity, $v_g$ [km/s] & 5.32$\pm$0.03 & 5.28$\pm$0.10 & 5.27$\pm$0.04 & 5.28$\pm$0.06 & 5.25$\pm$0.10 \\
        Central frequency, $f$ [GHz] & 0.89 & 0.97 & 0.83 & 0.94 & 0.74 \\
        Central wavevector, $k$ [rad/$\mu$m] & 1.0 & 1.1 & 0.9 & 1.1 & 0.85 \\
        Velocity dispersion slope along [100] $\frac{dv_p}{dk}$ [$\mu$m$^2$/rad/ns] & -0.21 & -- & -0.25 & -- & -0.23 \\
        Velocity dispersion slope along [110] $\frac{dv_p}{dk}$ [$\mu$m$^2$/rad/ns] & -0.29 & -- & -0.27 & -- & -0.31 \\
        In-plane anisotropy of phase velocity, $v_p^{[100]}/v_p^{[110]}$ & 0.972 & -- & 0.976 & -- & 0.969 \\
        In-plane anisotropy of group velocity, $v_g^{[100]}/v_g^{[110]}$ & 0.980 & -- & 0.985 & -- & 0.993 \\
        \hline
    \end{tabular}
    \label{tab:params}
\end{table}

\textcolor{new}{In Refs.~\cite{Ourdani_2024,ourdani_phon_mag_res_BLS} the high-frequency thermal SAWs were studied by BLS in FeRh/MgO(001) and the $f(k)$ dispersions were extracted for a range of wavevectors 25-140~rad/$\mu$m, i.e. above the one accessed in our experiments.
The FeRh thickness was higher (200~nm), and the SAW wavelengths were smaller than in our experiments, meaning that the FeRh film had more pronounced effect on the SAW parameters.
The value of $v_p$ for quasi-Rayleigh SAW was found to be larger for the FM phase of FeRh, which is consistent with our results (Figure~\ref{fig_speed}).
The dispersion relation obtained in Ref.~\cite{Ourdani_2024} gives approximately 4 times smaller frequencies for the same wavenumber than in our work.
This results from the fact that SAW dispersion scales with the product of the wavenumber and film thickness $kh$ \cite{FARNELL197235}, and the FeRh film in our study is four times thinner than that used in Ref.~\cite{Ourdani_2024}.
However, the range of SAW wavenumbers, which can be deliberately generated with the known techniques, was not explored in these works, and the SAW anisotropy was shown only for the FM FeRh phase and two principal crystallographic directions \cite{ourdani_phon_mag_res_BLS}.}

\textcolor{new}{In Refs.~\cite{SAW_FeRh_transition,FeRh_SAW_neuro_2025,FeRh_SAW_2026,SAW_in_FeRh_FMR} the CW SAWs with frequencies up to 0.9~GHz, similar to a central frequency in our study, were considered, generated electrically by antennae on a bare substrate and injected into the FeRh layer.
This limited the authors to piezoelectric substrates (GaAs or LiNbO$_3$) requiring buffer layers to achieve good quality of FeRh films, which affect the SAWs.
These differences hinder the direct comparison of the SAW velocities or dispersions.
The authors of Ref.~\cite{FeRh_SAW_2026} found an anomaly of $v_p\sim0.4$\% across the phase transition showing a hysteretic dependence on temperature.
However, such a small change in velocity cannot be seen in our data for laser-generated SAW pulses (Figure~\ref{fig_speed}) due to experimental uncertainties stemming from smaller propagation distance (up to 160~$\mu$m in our study vs 3500~$\mu$m) and 2-dimensional propagation as opposed to one-dimensional.
The $v_g(f)$ dispersion obtained in Ref.~\cite{FeRh_SAW_2026} is consistent with our results (Figure~\ref{fig_dispersion}) featuring a negative slope.
However, the chirp effect cannot be easily shown or exploited for a CW SAW as opposed to broadband pulses characterized in our work.
The anisotropy of SAW propagation also was not investigated since it is linked to a substrate material and would require a growth of additional antennae.
This brief consideration highlights that our results add information on SAW dispersion, anisotropy and chirping in FeRh/MgO structure in a wavevectors range highly relevant for picosecond acoustics.}

\section{Conclusions}

In conclusion, we have systematically investigated laser-generated surface acoustic wave pulses in a 60 nm epitaxial monocrystalline Fe$_{49}$Rh$_{51}$ thin film on a MgO(001) substrate using time-resolved Sagnac interferometry.
The photoinduced AFM-FM phase transition in FeRh was employed to modulate the efficiency of SAW generation, while the thermally induced transition allowed tuning of the thin-film acoustic load, thereby affecting the SAW propagation.
The SAW amplitude is strongly influenced by the excitation mechanism, whether thermoelastic or phase-transition-related, and exhibits nonlinear temperature and fluence dependencies, consistent with a previous report \cite{Previous}.
The range of generated SAW wavenumbers $k$=0.4-2~rad/$\mu$m with a central frequency of 0.9~GHz is governed by the laser spot size and remains independent of the excitation mechanism.
Phase and group velocities are primarily determined by the substrate.
A fourfold in-plane anisotropy of $v^{[100]}/v^{[110]}\approx$0.975 for the phase and 0.985 for the group velocity in both the AFM and FM phases is closer to unity than that in bare MgO.
The SAW phase velocity dispersion defined by the FeRh film manifests itself as a pronounced chirping of the pulse with higher phonon frequencies trailing in its tail.
The velocity dispersion quantified by the slope $dv_p/dk$ is higher along the MgO[110] (FeRh[100]) direction.
The velocities, their dispersions and anisotropies show negligible sensitivity to the phase of FeRh film, ensuring robustness of timings in a potential magnetoacoustic FeRh device, while the SAW amplitudes are strongly affected by the phase transition.
Such a combination of properties is important for FeRh-based neuromorphic computations conceptualized recently for continuous-wave SAWs \cite{FeRh_SAW_neuro_2025}.
Collectively, these results provide essential information for the design of optically controlled SAW-driven spintronic devices based on the AFM-FM transition in FeRh on MgO(001).

\medskip

\medskip
\textbf{Acknowledgements} \par 
Ia.~A.~M. acknowledges the support of the Russian Science Foundation grant No.~24-72-00111.

\medskip

%
\bibliographystyle{MSP}
\bibliography{FeRh_SAW_parameters}



\end{document}